\documentclass[]{article}
\newcommand{\bb}{\begin{eqnarray}}
\newcommand{\ee}{\end{eqnarray}}

\begin{document}
\title{Zeta function regularization, anomaly and complex mass term}
\author{P. Mitra\\
Saha Institute of Nuclear Physics, Calcutta 700064\\ 
parthasarathi.mitra@saha.ac.in}
\date{}
\maketitle
%
\begin{abstract}
If the zeta function regularization is used and 
a complex mass term  considered for fermions, the phase does 
not appear in the fermion determinant. 
This is not a drawback of the regularization,
which can recognize the phase through source terms, 
as demonstrated by the anomaly equation
which is explicitly derived here for a complex mass term.
\end{abstract}

\section{Introduction}

The predictions of a field theory can be formally expressed in terms of
functional integrals. For the fermionic sector, one writes
\bb
\int d\mu\exp[\int\bar\psi {\cal D}\psi]={\rm det} {\cal D},
\ee
where ${\cal D}$ stands for the appropriate Dirac operator for the case.
Apart from a kinetic part, the interaction with gauge
fields is also included. The mass term is often taken to be the simple
$-\int\bar\psi m\psi$, but the possibility of
CP violation requires the consideration of what is sometimes called a
twisted mass or simply a complex mass \cite{baluni}, namely 
$-\int\bar\psi m\exp (i{\theta}\gamma^5)\psi$. It is this $\theta$-term which
will be of central interest in the following. 

At the classical level, the phase $\theta$ may be removed by a 
chiral transformation, but in the full quantum field theory, the situation
is more complicated because of the chiral anomaly. It is well known now
that the fermion measure is not invariant under a chiral transformation,
so that an attempt to remove $\theta$ by such a transformation
may produce a non-trivial Jacobian dependent on $\theta$, thus causing the
reappearance of this parameter. However, the anomaly is a result of
short distance singularities, and needs to be studied with a proper
regularization. We shall discuss the consequences of the $\theta$-term
in the context of a specific regularization, the zeta function regularization.
This approach has been shown to
yield a functional integral independent of $\theta$, which
may suggest a limitation of the approach. However, the phase
has to appear in Green functions and it duly does,
when these are calculated by using sources.
The anomaly equation too
has to contain $\theta$, not in the anomaly but in the classical
mass term. It is not obvious that this will occur in the zeta function 
framework which makes use of the $\theta$-independent Laplacian operator.
Hence we study the anomaly equation in 
the case of a complex mass term and verify that the regularization is 
capable of accommodating the phase as well as the anomaly.
This demonstrates the reliability of the
zeta function regularization and makes the use of this approach to the
regularization of the determinant acceptable even in this case.
\section{Review of fermion determinant in zeta function regularization}

The determinant of a matrix can be thought of as the
product of its eigenvalues. For an operator, the
product of the eigenvalues has to be regularized.

The zeta function regularization is widely used in mathematical discussions 
in quantum field theory \cite{dowker, hawk, zeta}. It was shown quite a while 
back that the chiral anomaly in vector gauge theories can be evaluated by 
using this regularization without recourse to Feynman diagrams \cite{reuter}. 

The zeta function of an operator $X$ involves a parameter $s$,
\bb
\zeta(s,X)\equiv{\rm Tr}(X^{-s}).
\ee
In terms of eigenvalues $\lambda$, this becomes
\bb
\zeta(s,X)=\sum(\lambda^{-s}),
\ee
so that
\bb
\zeta'(s,X)=-\sum(\ln\lambda \lambda^{-s}),
\ee
and
\bb
\zeta'(0,X)=-\sum(\ln\lambda )=-\ln\prod\lambda=-\ln{\rm det} X.
\ee
This provides a definition of the determinant.
The eigenvalues are assumed to be positive in this definition.

Note that the Dirac operator
\bb
{\cal D}=i\not{\!\! D}-m\exp (i{\theta}\gamma^5)
\ee
is neither hermitian nor antihermitian even for a real mass term. 
A {\it positive} operator is constructed for the zeta function by
going over to the Laplacian from the Dirac operator as in \cite{reuter}: 
\bb
\Delta=[i\not{\!\! D}-m\exp (i{\theta}\gamma^5)]^\dagger
[i\not{\!\! D}-m\exp (i{\theta}\gamma^5)].
\ee
Formally, the two factors are related by chiral transformations
and formally the determinants of all these transformations can be taken to
be unity,
so the determinant may be said to have been squared in the process,
and a square root has to be included in the definition of the determinant.

For {\it antihermitian} $\gamma$-matrices, as appropriate for
euclidean spacetime, $\Delta$ is independent of the phase ${\theta}$:
\bb
\Delta=(\not{\!\! D})^2+m^2.
\ee
The zeta function of this operator is
\bb
\zeta(s,\Delta)\equiv{\rm Tr}(\Delta^{-s}),
\ee
and the regularized logarithm of the functional integral 
is defined in the limit of $s\to0$ as
\bb
\ln Z\equiv-\frac12\zeta'(0,\Delta)-\frac12\ln\mu^2\zeta(0,\Delta).
\ee
The square root is introduced because of the squaring in the
construction of $\Delta$ mentioned above.
It is to be noted that the determinant is defined only for the
product $\Delta$ and not for the Dirac operators.

The regularized determinant is independent of ${\theta}$, 
depending on the gauge fields
only through the operator $\Delta$ and is therefore
invariant under symmetry transformations of the gauge field $A$ \cite{PM}.

This may be compared with the formal determinant of a Dirac operator when
$\not{\!\! D}$ has only a finite number of zero modes and no other eigenvalue.
Because of the anticommutation of $\not{\!\! D}$ and $\gamma^5$, the zero modes
can be chosen to be of definite chirality. The mass term produces a factor of 
$\exp (i{\theta}\gamma^5)$
for each zero mode, leading to a product $\exp (i{\theta}\nu)$ where $\nu$
stands for the number of positive chirality zero modes {\it reduced} by the
number of negative chirality zero modes. This number depends on the gauge field
involved in $D_\mu$. Such a factor will continue to appear if a finite number
of nonzero modes occur. However, when the number of modes becomes infinite,
the reordering of the eigenvalues involved in identifying such a factor
is not admissible and regularization is crucial.
A regularization may even remove the $\theta$ dependence. But
it has to be checked whether the regularization vitiates the anomaly equation.
\section{Inclusion of fermion sources}
If one wants to calculate fermion Green functions, one has to introduce
fermion source terms in the standard way. This means the consideration of
\bb
\int d\mu\exp[\int(\bar\psi {\cal D}\psi +\bar\psi\eta+\bar\eta\psi)]=
{\rm det} {\cal D}\exp[-\int\bar\eta{\cal D}^{-1}\eta].
\ee
The determinant is defined by the zeta function method indicated
above, while the source dependent factor is separate. This factor
explicitly involves $\theta$ through ${\cal D}$. Thus fermionic Green
functions continue to depend on $\theta$ in the zeta function approach. This is
because the fermion field is not chiral invariant. 
There is no contribution to the $\theta$ dependence from fermion loops in the
determinant.
As an example, the propagator is given by
\bb
\langle\psi(x)\bar\psi(y)\rangle=\langle{\cal D}^{-1}(x,y)\rangle,
\ee
where the averaging is over gauge fields with the effective gauge field
action arising from the original gauge field action and including the
effect of the fermion determinant. This depends on $\theta$ through
${\cal D}^{-1}$.
The only $\theta$ dependence however goes away when the external legs
in a Feynman diagram
are amputated, {\it i.e.,} the ${\cal D}^{-1}$-s are removed by ${\cal D}$.
\section{Anomaly equation and $\theta$}

The $\theta$-independence of the determinant
may suggest that the zeta function approach,
relying as it does on the product of the Dirac operator with its conjugate,
is somewhat handicapped and not sensitive to the presence of $\theta$.
If this were the case, it would be a serious problem for the 
zeta function approach.
It has already been pointed out that Green functions do contain $\theta$.
The crucial anomaly equation also has to involve the parameter $\theta$:
\bb
\partial^\mu\langle \bar\psi\gamma_\mu\gamma_5\psi\rangle=
2im\langle \bar\psi\gamma_5\exp (i{\theta}\gamma^5)\psi\rangle
+{\rm anomaly}.
\nonumber\ee
This is particularly significant because the possibility of obstructions
to the removal of $\theta$ from the fermion action arises entirely
from the anomaly.
Although the anomaly has been demonstrated in this approach, it was for
a real mass term, where  $\theta=0$. The anomaly is supposed to be
independent of the mass, so one can expect it to arise also for a
complex mass term, but does $\theta$ appear as indicated in the zeta
function approach?

To derive the anomaly equation in this framework,
one has to add source terms for the composite fermion operators
in that equation to the Dirac operator \cite{reuter}.
The phase ${\theta}$ requires us to consider the modified operator  
\bb
[i\not{\!\! D}-i\not{\!\! Q}\gamma^5-m\exp (i{\theta}\gamma^5)
-K\gamma^5\exp (i{\theta}\gamma^5)],
\ee
with $Q^\mu(x)$ coupling to the axial current and $K$
to the pseudoscalar density including the phase. 
Note that in the presence of the phase in the mass term, the parity
symmetry transformation of the action is chirally rotated, so
that $\bar\psi\exp (i{\theta}\gamma^5)\psi$ is a scalar and
$\bar\psi\gamma^5\exp (i{\theta}\gamma^5)\psi$ is a pseudoscalar.
That is why the pseudoscalar source has to have the chiral phase factor.
This leads to the modified Laplacian
\bb
\Delta'&=&[i\not{\!\! D}-i\not{\!\! Q}\gamma^5-m\exp (i{\theta}\gamma^5)
-K\gamma^5\exp (i{\theta}\gamma^5)]^\dagger\nonumber\\
&&[i\not{\!\! D}-i\not{\!\! Q}\gamma^5-m\exp (i{\theta}\gamma^5)
-K\gamma^5\exp (i{\theta}\gamma^5)]\nonumber\\
&=& (\not{\!\! D})^2 +m^2 +K^2-(\not{\!\! Q})^2-\not{\!\! Q}\gamma^5
\not{\!\! D} -\not{\!\! D}\not{\!\! Q}\gamma^5+2mK\gamma^5\nonumber\\
&& +i(\not{\!\! D}-\not{\!\! Q}\gamma^5)K\gamma^5\exp (i{\theta}\gamma^5)
-iK\gamma^5\exp (-i{\theta}\gamma^5)(\not{\!\! D}-\not{\!\! Q}\gamma^5),
\ee
which does depend on $\theta$.
This $\Delta'$ operator is used to define a modified $Z'$ and thence
the expectation value of the axial current operator
\bb
\langle \bar\psi\gamma_\mu\gamma_5\psi\rangle=
i{\delta\ln Z'\over\delta Q^\mu(x)}|_{Q=K=0}=
-\frac{i}{2}{\delta\zeta'(0,\Delta')\over\delta Q^\mu(x)}|_{Q=K=0}.
\ee
Now let $\phi_n$ be eigenfunctions and $\lambda_n$ the 
corresponding eigenvalues for $\Delta$, primed as required, but see below:
\bb
\Delta\phi_n=\lambda_n\phi_n, \quad \Delta'\phi_n'=\lambda_n'\phi_n'.
\ee
Then by the Hellman - Feynman theorem,
\bb
{\delta\zeta(s,\Delta')\over\delta Q^\mu(x)}|_{Q=K=0}
&=&
\sum_n{\delta\lambda'^{-s}_n\over\delta Q^\mu(x)}|_{Q=K=0}\nonumber\\
&=&-s\sum_n\lambda^{-s-1}_n{\delta\lambda'_n\over\delta Q^\mu(x)}|_{Q=K=0}
\nonumber\\
&=&-s\sum_n\lambda^{-s-1}_n\int d^4w\phi^\dagger_n(w){\delta\Delta'\over
\delta Q^\mu(x)}|_{Q=K=0}\phi_n(w)\nonumber\\
&=&s\sum_n\lambda^{-s-1}_n\phi^\dagger_n(x)[\gamma_\mu\gamma^5\not{\!\! D}
-\stackrel{\leftarrow}{\not{\!\! D}}\gamma_\mu\gamma^5]\phi_n(x).
\ee
On taking the divergence 
one can simplify the expression if one takes the $\phi_n$
to be eigenfunctions of $\exp (-i{\theta}\gamma^5)\not{\!\! D}$ in addition to 
$\Delta=(\not{\!\! D})^2+m^2$. 
\bb
\exp (-i{\theta}\gamma^5)\not{\!\! D}\phi_n=\alpha_n\phi_n.
\ee
This is possible with 
\bb
\alpha_n^2+m^2=\lambda_n
\ee
because 
$\exp (-i{\theta}\gamma^5)\not{\!\! D}$ is hermitian and its square is the same 
as $(\not{\!\! D})^2$, which differs only by $m^2$ from $\Delta$. The 
exponential factor is not relevant here, it just cancels out on squaring the 
operator and does not cause any problem; however, it is needed at a later 
stage. One finds
\bb
\partial^\mu\langle \bar\psi\gamma_\mu\gamma_5\psi\rangle&=&
2i\sum_n[s\lambda^{-s}_n\phi^\dagger_n(x)\gamma^5\phi_n(x)]'|_{s\to 0}
\nonumber\\
&&-2im^2\sum_n[s\lambda^{-s-1}_n\phi^\dagger_n(x)\gamma^5\phi_n(x)]'|_{s\to 0}.
\label{div}
\ee
The first term of (\ref{div}) is reminiscent of
$2i\sum_n\exp(-\lambda_n/M^2)\phi^\dagger_n(x)\gamma^5\phi_n(x)$ 
in the measure approach and similar calculations \cite{reuter} show it to be 
the anomaly term
\bb
\frac{i}{16\pi^2}{\rm tr}\epsilon^{\mu\nu\rho\sigma}
F_{\mu\nu}F_{\rho\sigma}.
\ee
The second term of (\ref{div}) does not depend explicitly on the phase $\theta$
but the eigenfunctions $\phi_n$ implicitly involve it. To rewrite the term
in a more familiar form, one has to recognize
\bb
\langle \bar\psi\gamma_5\exp (i{\theta}\gamma^5)\psi\rangle=
-{\delta\ln Z'\over\delta K(x)}|_{Q=K=0}=
\frac{1}{2}{\delta\zeta'(0,\Delta')\over\delta K(x)}|_{Q=K=0}.
\ee
Now as in the case of the axial vector above, one has
\bb
{\delta\zeta(s,\Delta')\over\delta K(x)}|_{Q=K=0}
&=&
\sum_n{\delta\lambda'^{-s}_n\over\delta K(x)}|_{Q=K=0}\nonumber\\
&=&-s\sum_n\lambda^{-s-1}_n{\delta\lambda'_n\over\delta K(x)}|_{Q=K=0}
\nonumber\\
&=&-s\sum_n\lambda^{-s-1}_n\int d^4w\phi^\dagger_n(w){\delta\Delta'\over
\delta K(x)}|_{Q=K=0}\phi_n(w)\nonumber\\
&=&-s\sum_n\lambda^{-s-1}_n\phi^\dagger_n(x)[-i\gamma^5\exp (-i{\theta}\gamma^5)
\not{\!\! D}\nonumber\\
&& -i\stackrel{\leftarrow}{\not{\!\! D}}\gamma^5\exp (i{\theta}\gamma^5)
+2m\gamma^5]\phi_n(x).
\ee
This may be simplified by using the fact that $\phi_n(x)$ is an eigenfunction of
the operator $\exp (-i{\theta}\gamma^5)\not{\!\! D}$ in addition to 
$\Delta=(\not{\!\! D})^2+m^2$, as mentioned earlier. One finds
\bb
\langle \bar\psi\gamma_5\exp (i{\theta}\gamma^5)\psi\rangle=
-m\sum_n[s\lambda^{-s-1}_n\phi^\dagger_n(x)\gamma^5\phi_n(x)]'|_{s\to 0},
\ee
so that
\bb
\partial^\mu\langle \bar\psi\gamma_\mu\gamma_5\psi\rangle=
2im\langle \bar\psi\gamma_5\exp (i{\theta}\gamma^5)\psi\rangle
+\frac{i}{16\pi^2}{\rm tr}\epsilon^{\mu\nu\rho\sigma}
F_{\mu\nu}F_{\rho\sigma}.
\ee
This is the anomaly equation in euclidean space and it contains $\theta$ as 
expected in the mass term.  This confirms that the zeta function approach 
is not blind to $\theta$ even though the determinant is.

\section{Conclusion}

To conclude, we have reproduced the anomaly equation for a complex mass term 
using the zeta function approach. 
The anomaly assures us that the $\theta$
independence of the determinant in the zeta function approach
is not due to any inability to perceive the anomaly.
Apart from the anomaly, there is the non-anomalous mass-dependent
piece in the equation which has to contain the phase when the mass term in the
action contains it. This is what has been checked by generalizing
the derivation for real mass terms.
In doing so, the pseudoscalar source and
the eigenfunctions have had to be reorganized because of the phase. The 
importance of the calculation is that it shows that the 
use of the Laplacian operator, which is
{\it independent of the phase}, does not compromise 
the power of the zeta function approach and thus makes this
approach acceptable even in the context of complex mass terms.

\end{document}